\documentclass[twocolumn,notitlepage,showpacs,prl,superscriptaddress]{revtex4-1}

\usepackage{amsmath}
\usepackage{epsf}
\usepackage{graphicx}
\usepackage{dcolumn}
\usepackage{bm}

\newcommand{\req}[1]{Eq.~(\ref{#1})}
\newcommand{\reqs}[1]{Eqs.~(\ref{#1})}
\newcommand{\rref}[1]{(\ref{#1})}

\newcommand{\p}{\mathbf{p}}

\renewcommand{\r}{\mathbf{r}}

\newcommand{\beq}{\begin{equation}}
\newcommand{\eeq}{\end{equation}}
\newcommand{\be}{\begin{equation}}
\newcommand{\ee}{\end{equation}}
\newcommand{\beqa}{\begin{eqnarray}}
\newcommand{\eeqa}{\end{eqnarray}}
\newcommand{\bea}{\begin{align}}
\newcommand{\eea}{\end{align}}
\usepackage{amssymb}
\usepackage{array}
\usepackage{amsmath}
\usepackage{graphicx}\usepackage{graphics}
\usepackage{dcolumn}
\usepackage{bm}\usepackage{varioref}

\usepackage{amsmath}
\usepackage{epsf}
\usepackage{graphicx}

\usepackage{subfigure}
\usepackage{dcolumn}
\usepackage{bm}

\usepackage{amssymb}
\usepackage{array}
\usepackage{varioref}
\usepackage{wrapfig}
\usepackage{etoolbox}
\usepackage{color}

\providecommand{\bv}[1]{\bm{\mathrm{#1}}}

\providecommand{\p}{\bv{p}}

\renewcommand{\r}{\bv{r}}

\providecommand{\zb}{{\bar z}}
\renewcommand{\b}[1]{\bar #1}
\providecommand{\cc}{\mbox{c.c.}}

\makeatletter
\newcommand*{\balancecolsandclearpage}{%
  \close@column@grid
  \clearpage
  \twocolumngrid
}
\makeatother

\begin{document}

\title{
 Instability of Abrikosov lattice due to nonanalytic core reconstruction of vortices in Bosonic superfluids}

\author{Avraham Klein}
\affiliation{The Racah Institute of Physics, The Hebrew University of Jerusalem, 91904, Israel}
\author{Oded Agam}
\affiliation{The Racah Institute of Physics, The Hebrew University of Jerusalem, 91904, Israel}
\author{Igor L. Aleiner}
\affiliation{Physics Department, Columbia University, New York, NY 10027, USA}

\date{\today}
\begin{abstract}
We study the impact of the non-analytic reconstruction of vortex cores  on static vortex structures in weakly coupled superfluids. We show that in rotating two-dimensional systems, the Abrikosov vortex lattice is unstable to vortex core deformation: Each zero of the wave function becomes a cut of finite length. The directors characterizing the orientations of the cuts are themselves ordered in  superstructures due either to surface effects or to interaction with shear deformations of the lattice (spiral structure). Similar instability may be also observable in clean superconducting films.

\end{abstract}

\pacs{67.10.-j, 67.10.Jn,67.85.-d ,67.85.De}

\maketitle

{\em Introduction} -- A rapidly rotating bosonic superfluid forms into a dense lattice of vortices rotating as a rigid body - the Abrikosov lattice \cite{Abrikosov1957a,Abrikosov1957}. The  favorable lattice is triangular, and its stability and vibrational modes were first studied by Tkachenko \cite{Tkachenko1966,Tkachenko1966a}. Tkachenko waves were first detected in superfluid helium \cite{Andereck1980,Andereck1982}, and later were observed directly in rotating cold atomic condensates \cite{Coddington2003}.

The triangular lattice structure minimizes the energy of the 
currents created by the quantized vortices in the rotating frame. 
Alternatively, using the well-known analogy between quantized vortices in two-dimensional (2D) superfluids and point charges in 2D electrodynamics \cite{Popov1973,Popov2001,Ambegaokar1980}, it is the minimum energy configuration of point charges in a neutralizing uniform background field. However, the point-charge approach does not take into account the possibility of core deformation in the vortices, which endows each vortex with an additional dipole-like degree of freedom.

In the strong coupling regime, namely when core size of the vortex is of order to the interparticle distance, this core deformation is negligible. However, Bose-Einstein Condensates 
 of cold atoms may be  realized in the weak coupling limit,  were core size is much larger than the interparticle distance. 
Here, it has been recently shown that vortices 
in 2D bosonic superfluids experience non-analytic reconstruction of their cores when moving with respect to the ambient superfluid \cite{Klein2014}. In this Letter we show that this core reconstruction leads to an instability of the Abrikosov lattice, and identify the new ordering.

{\em Non-analytic reconstruction of the vortex core} --
To begin with, let us summarize the relevant findings of the analytic theory \cite{Klein2014} of vortices in superfluids
at the weak coupling regime, $n_0\xi^2 \gg 1$, where $n_0$ is the 2D bosonic density and $\xi$ is the healing length:
i) The low energy dynamics are not exhausted by the position of the vortex
attached to the ambient flow but must include also the motion of the vortex with respect to this flow;
ii) This motion can be characterized by the kinetic momenta $\hat{\p}=(\hat{p}_{x},\hat{p}_y)$;
iii) The momentum dependence of the kinetic energy of the vortex circular motion is non-analytic
\be
{H}({\p})=\frac{{\p}^2}{2M_v({\p}^2)};
\  \frac{M_v(p^2)}{m n\xi^2}\equiv
\frac{\pi}{\alpha^2} \ln\left(\frac{\hbar^2 n^2\xi^2}{p^2} \right),
\label{eq:1b}
\ee
where $m$ is the boson mass, and $\alpha=0.8024\dots$. The non-analytic dependence of the {\em added mass}, $M_v$, on the momentum $p$ is associated with the weak solutions of the Gross-Pitaevskii equation inside the vortex core \cite{Klein2014},
see also \cite{Radzihovsky15} for singular vortex antivortex configurations.

An effective theory\cite{Klein2014} describing a superfluid system with  $N$ vortices is conveniently written
using Popov's formalism \cite{Popov2001} which maps 2D superfluidity to 2D nonlinear electrodynamics. In this mapping, vortices become charged particles characterized by the coordinates ${\bf R}_k(t)$, and charge $2\pi\hbar\sigma_k $ with $\sigma_k=\pm 1$ being the vorticity. The electric field ${\bf E}$ is related to the superfluid current ${\bf j}$ as ${\bf E}=-\hat{\epsilon}{\bf j}$ (where $\hat{\epsilon}$ is the antisymmetric tensor of the second rank acting on the spatial
coordinates), and the magnetic field, $B$, is the boson density, $n$.
The physical fields ${\bf E},B$ are related to the vector ${\bf A}$ and scalar $\varphi$ potentials by usual relations
${\bf E}=-\partial_t {\bf A}-\nabla \varphi$, and $B=\nabla\times {\bf A}$.
In these variables the Lagrangian ${\cal L}$ (describing the system over linear scales much larger than the healing length $\xi$) of a rotating superfluid with angular frequency $\Omega$ is
\be
\begin{split}
&
{\cal L}=\sum_{j=1}^N{\cal L}_v\left({\bf p}_j;\,\sigma_j;\,\dot{\bf R}_j;\,
\left\{
{\bf E};{B};{\bf A}; {\varphi}
\right\}_{\r={\bf R}_j}
\right)+\int d^2r{\cal L}_f;
\\
&
{\cal L}_v= {\bf p}\cdot \left(\dot{\bf R}
-\hat{\epsilon}\frac{{\bf E}}{B}
\right)+2\pi\hbar\sigma\left({\bf A} \cdot \dot{\bf R}-\varphi \right)  - {H}\left(\p\right);
\\
&
{\cal L}_f=\frac{m}{2}\left[\frac{\left({\bf E}
+\Omega B\r\right)^2}{B}-\frac{c^2\left(B-n_0\right)^2}{n_0}\right],
\end{split}
\raisetag{1cm}
\label{Lagrangian}
\ee
where $c\equiv \hbar/(m\xi)$ is the sound velocity.
Equation \rref{Lagrangian} is dictated by the Galilean invariance of the system and the transformation to a rotating frame. Its gauge invariance is the vorticity conservation. The corresponding action $\int {\cal L}dt$ should minimized subject to
the condition $\varphi=0$ at the boundary of the 2D superfluid. (The physical meaning of this condition is that the energy 
of current generated due to placing a vortex in the system is counted from the energy of the vortex on the boundary.)

The $\hat{\epsilon}\frac{{\bf E}}{B}$ term in ${\cal L}_v$ deserves some discussion. In the first place, it highlights the fact that the spectrum of the $H(p)$ is related to the reconstruction of the core of the vortex moving with respect to ambient flow and thus should vanish for the vortex moving together with the flow,  $\dot{\bf R}=
\hat{\epsilon}{{\bf E}}/{B}$. Next, unlike charged particles in vacuum, 
the vortex can be at rest even for $\p\neq 0$ provided that ambient flow is finite. Finally, any reconstruction of the core
 gives rise to an emergent dipole moment $2\pi\sigma_j \hbar {\bf d}_j$, with the displacement vector
 \be
{\bf d}_j=  \sigma_j\frac{ \hat{\epsilon} {\bf p}_j}{2\pi\hbar B}.
\ee

{\em Energy of an array of vortices with deformed cores} -- In what follows we assume that $N$ vortices are sufficiently 
far apart so that one may assume the density to be constant, $B=n_0$. For simplicity, we assume that the $2D$ superfluid is
confined in a circle of radius $ R\gg \xi$, and neglect the effect of image vortices, whose main effect is to reduce the effective radius of the system by order of $a\sqrt{\ln(a/\xi)}$, where $a$ is the lattice constant \cite{Stauffer1968} . Then, looking for a stationary solution $\dot{\bf A}=0;\ \dot{\bf R}=0$, we minimze the action with respect to $\varphi$ and obtain the energy
of the vortices (with $\sigma_j=\mbox{sgn} \Omega$) as a function of their positions and displacement vectors:
\be
\begin{split}
&U=\frac{\pi\hbar^2n_0}{m}
\left[
N^2\ln\frac{ R}{\xi}
-N\frac{ m|\Omega|R^2}{\hbar}+u\right];
\\
u&=\sum_{j=1}^N  \left[\varepsilon\left(\frac{{\bf d}_j^2}{\xi^2}\right)
+ \frac{ m|\Omega|}{\hbar} \left( {\bf R}_j^2 + 
2 {\bf d}_j\cdot {\bf R}_j \right)\right]
\\
&
 +  \sum_{i\neq j}^N  \left(1+ {\bf d}_i 
\cdot \frac{\partial}{\partial {\bf R}_i}\right) \left(1+ 
{\bf d}_{j} \cdot \frac{\partial}{\partial {\bf R}_{j}}\right)\ln \frac{\xi}{|{\bf R}_i-{\bf R}_{j}|},
\end{split}
\raisetag{2cm}
\label{eq:3}
\ee
with the dimensionless function $\epsilon(x)$  given by
\be
\varepsilon(x)=\frac{4\alpha^2 x}{\ln(1/x)}. \label{epsilon}
\ee
For ${\bf d}_j=0$, \req{eq:3} reduces to the well known energy of logarithmically repelling particles in a parabolic confinement
potential, yielding the configuration of an Abrikosov  lattice with non-deformed cores. Minimization of the energy \rref{eq:3} 
with respect to the vortex positions ${\bf R}_i$ gives
 \be
\frac{ \partial}{\partial {\bf R}_i} \left[ \sum_{j\neq i}^N \ln\frac{1}{|{\bf R}_i-{\bf R}_{j}|}
+ \frac{m|\Omega|}{\hbar} \sum_j^N {\bf R}_{j}^2 \right] =0.
\label{minimum1}
\ee
The solution of this equation, ${\bf R}_j={\bf r}_j$, produces the sites of the lattice corresponding to local energy minima. It is known that the global minimum corresponds to ${\r}_j$ comprising a triangular lattice \cite{Tkachenko1966a}.
One useful relation can be derived independently of any lattice structure.
Multiplying both sides of \req{minimum1} by  ${\r}_i$ and summing over $i$, we obtain
 \be
\frac{m |\Omega|}{\hbar} \sum_j^N {\bf  r}_{j}^2 = \frac{N(N-1)}{2},
\label{Omega} 
\ee
thus, vortices start to overlap at $\Omega\geq \Omega_c= 2\pi \hbar/(\sqrt{3} m \xi^2)$ for the triangular lattice.
For $\Omega<\Omega_c$, the distance between nearest vortices is
\be
a= \left(\Omega_c/|\Omega|\right)^{1/2}\xi.
\ee

{\em Instability of the Abrikosov lattice due to core deformation} -- Consider now a shift of the vortices with respect to their original position, ${\bf r}_j\to {\bf r}_j + \boldsymbol{\delta}_j$.
It will be convenient to work with the variable $\boldsymbol{\Delta}_j=\boldsymbol{\delta}_j+ {\bf d}_j$. The variable ${\bf r}_j+\Delta_j$ should be understood as the {\em virtual position} of the vortices, defined by the current distribution they create at distances larger than $\xi$.
Expanding the energy in \req{eq:3} to second order in ${\bf d}_j$ and $\boldsymbol{\Delta}_j$ we obtain that the quadratic contribution to $u$  is
 \be
\begin{split}
u_2&=u_2^\Delta+u_2^D;
\\
u_2^\Delta &=  \frac{m |\Omega|}{\hbar} \sum_i  |\Delta_i|^2+\frac{1}{2} \mbox{Re}
\sum_{i\neq j}^N \frac{ (\Delta_i-\Delta_j)^2} {(z_i-z_j)^2}
;
\\
 u_2^D &=   \sum_i \left[ \varepsilon\left(\frac{|D_i|}{\xi^2}\right) -\frac{m| \Omega|}{\hbar}|D_i|
- \mbox{Re}\, D_i\bar{V}_i^{(2)}
 \right];\
\\
\end{split}
\label{u2}
 \ee
 Here and henceforth we employ complex vector notation, i.e. $z=r_x+i r_y$, $d= d_x + i d_y$, $\Delta= \Delta_x + i \Delta_y$, the director is defined as $D_i\equiv d_i^2$, and the non-diagonal component of the electric field gradient $\bar{V}_i^{(2)}$ tensor is given by
\be
\bar{V}_i^{(n)} =  \sum_{j\neq i}^N \frac{1}{(z_i-z_j)^n}.
\label{Vi}
\ee
The $u_2^\Delta$ term describes the stability of the triangular lattice with respect to small displacements of the
vortex centers and it was studied
by Tkachenko \cite{Tkachenko1966}. The $u_2^D$ term describes the sensitivity of the Abrikosov lattice to core deformation and
we  turn to studying this term.

We notice that in the quadratic approximation \rref{u2} the  cores deformations are independent. This is because
each core deformation ${\bf d}_i$ is followed by the displacement of the vortex center by $\boldsymbol{ \delta}_i=-{\bf d}_i$ so that
the currents created by the vortex at the distance larger than the healing length $\xi$ remain intact.
Moreover, far from  the edges of the sample ${\bar V}_i^{(2)}\to 0$  due to the $C_3$ rotational symmetry.
Minimization with respect to $|D_i|$ with the logarithmic accuracy gives the absolute value of the director
\be
|d_i|^2=|D_i|=D_*= \xi^2 e^{- \gamma \Omega_c/|\Omega|},\
\label{D*}
\ee
where $\gamma={2\sqrt{3} \alpha^2 }/{\pi}\approx 0.74$.
The quantization of the vortex motion, \cite{Klein2014}, sets a lower bound on the value of the displacement vector $|D|n_0>1$.
Thus, \req{D*} is applicable in the frequency interval
\be
\Omega_c/\ln(n_0\xi^2) <\Omega <\Omega_c,
\label{condition}
\ee
{\em i.e.} it occurs only in the weak coupling regime, $n_0\xi^2\gg 1$.

Equation \rref{D*} predicting the size of the non-analytic deformation of the vortex cores in the Abrikosov lattice is the main result of this Letter. However, it implies local rotational symmetry with respect to the direction of the deformation. In what follows we study two groups of the effects lifting this symmetry: (i) boundary effects; and (ii) non-linear couplings.

\begin{figure}[t]
\includegraphics[width=0.9\columnwidth]{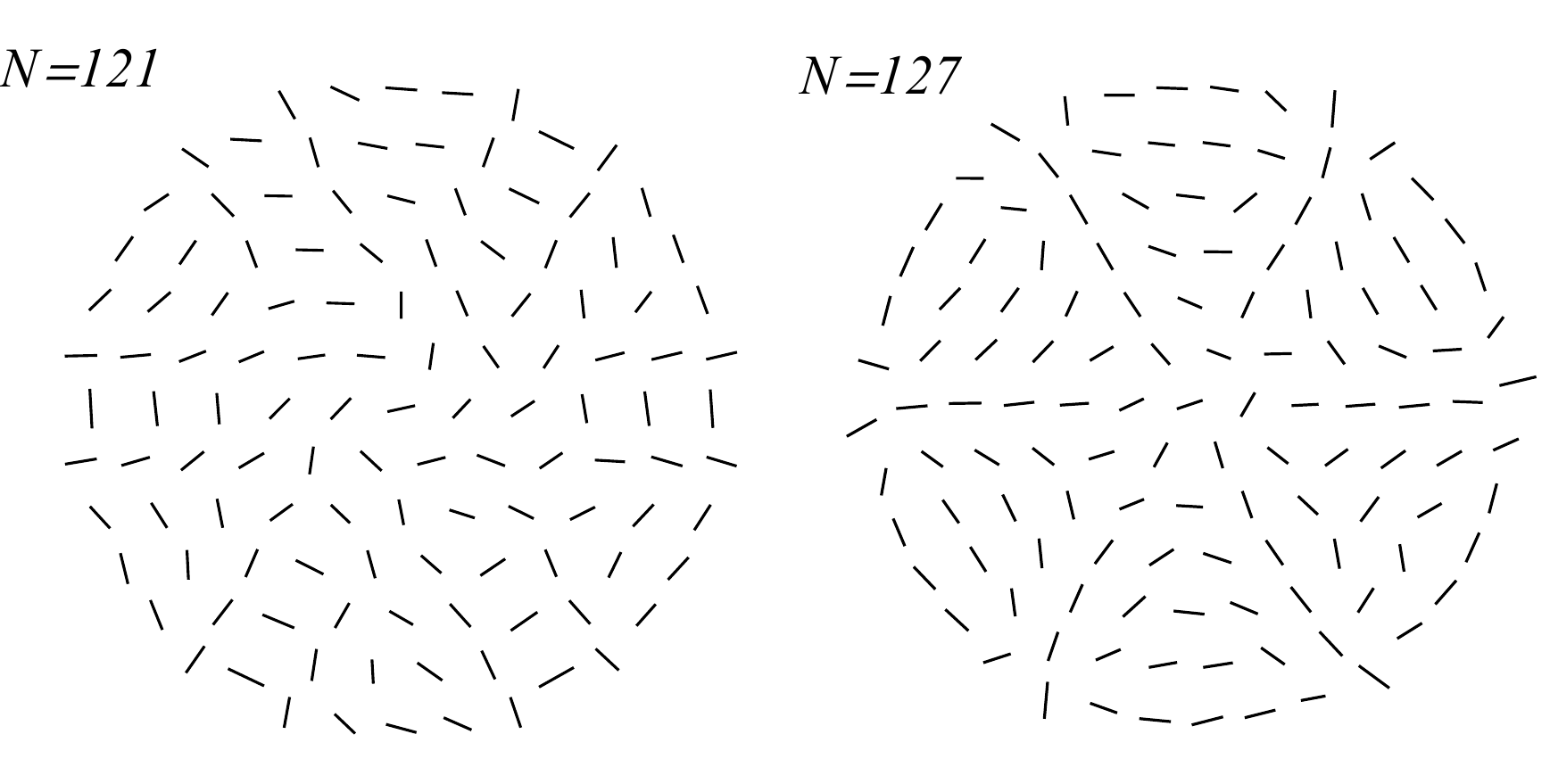}
 \caption{Deformation of a finite Abrikosov lattice due to core reconstruction [14]. The panels show a the configuration of the Abrikosov lattice obtained by numerical minimization of the energy $u$ for a system with 121(left) and 127 (right) vortices, taking into account their emergent dipole moments $d_i$. The lines in the figures represent the directors configuration: The length of each line is $10\sqrt{|D_i|}$, and each line has an angle corresponding to $(\arg\, D_i)/2$ in accordance with the convention for liquid crystals. The lattice constant is $a = 3$.}
 \label{fig1}
\end{figure}

 {\em Boundary effects}-- Consider a finite size lattice with a vortex at $z=0$ and $|z_j|<R$.  Vortex repulsion leads to lattice deformations near the boundary. Far away from the boundary, $|z_j| \ll R$, these deformations may be neglected \cite{Koulakov1998}, and performing finite sum in \req{Vi} we find \cite{Note}
\begin{subequations}
\be
\bar{V}^{(2)}_i\simeq \frac{1}{a^{1/2}R^{3/2}}
\left(\frac{{z}_i}{R}\right)^4 f\left(\frac{R}{a}\right),
\label{Vasymptoticsb}
\ee
 where $f(x)$ is an oscillatory function which may have either positive or negative sign \cite{Note}. Near the boundary, the value of the field gradient \rref{Vi} can be estimated using a straight boundary approximation,
\be
\bar{V}^{(2)}_i=
\frac{1}{a^2}\left(\frac{\bar{z}}{z}\right)\eta\left(\frac{z}{\zb}\right); ~~~~R-|z_i| < a
\label{eq:V2-surf}
\ee
where $\eta(x)$ is of order unity and strongly depends on the boundary facets \cite{Note}. 
It is noteworthy that the angular dependence in \reqs{Vasymptotics} is consistent with the $C_6^v$ symmetry of the underlying lattice.
\label{Vasymptotics}
\end{subequations}
Substituting \reqs{Vasymptotics} 
into \req{u2} and minimizing with respect 
to the angle between $D_i$ and cystallographic direction we find that 
\be
\frac{D_i}{|D_i|}=\mbox{sign}\left[f\left(\frac{R}{a}\right)\right] \left(\frac{\bar{z}_i}{|{z}_i|}\right)^4.
\label{supervortex1}
\ee
This behavior was verified by direct numerical minimization of $u$,
given by \req{eq:3} (see Fig.~\ref{fig1}).

It is clear from \req{Vasymptoticsb} that the characteristic anisotropy
energy $u_{an}=D_*\left|\bar{V}^{(2)}_i\right|$ diminishes with the increasing size of the system. Therefore, it is imperative to consider higher order nonlocal terms in $D_i$, as they establish long range order.

{\em Higher order corrections} to the quadratic energy \rref{u2} are obtained from  further expansion of \rref{eq:3} in $d_i, \Delta_i$. The results are most easily written in terms of the Nambu spinors 
\begin{subequations}\label{eq:u2u3u4}
\label{higherorder}
\be
{\mathbb R}_i=\begin{pmatrix}\Delta_i\\\bar{\Delta}_i\end{pmatrix};
\quad{\mathbb D}_i=\begin{pmatrix}D_i\\\bar{D}_i\end{pmatrix};
\quad
\begin{matrix}
\bar{\mathbb R}_i= \left(\bar{\Delta}_i,\,\Delta_i\right)\\
\bar{\mathbb D}_i= \left( \bar{D}_i,\,{D}_i\right)
\end{matrix}
\ee
where the complex notation was introduced after \req{u2}. The correction $\delta u=u_2^\Delta+u_3+u_4$ takes the form
\begin{align}
u_2^\Delta &=\frac{1}{2}\bar{\mathbb R}_i 
\hat{M}^{(2)}_{ij}{\mathbb R}_j \\
u_3&=-\bar{\mathbb R}_i\hat{M}^{(3)}_{ij}{\mathbb D}_j\\ u_4&=\frac{3}{4}\bar{\mathbb D}_i
\hat{M}^{(4)}_{ij}{\mathbb D}_j,
\end{align}
and we used Einstein summation over repeated indices.
The matrices $\hat{M}^{(n)}_{ij},\ n=2,3,4$ are defined as
\be
\hat{M}^{(n)}_{ij}
=\lambda\delta_{ij}\delta_{n,2}\hat{\openone}-
(1-\delta_{ij})
\begin{pmatrix}0; \quad \frac{1}{\left(\bar{z}_i-\bar{z}_j\right)^n}
\\
\frac{1}{\left({z}_i-{z}_j\right)^n}; \quad 0
\end{pmatrix},
\ee
and $\lambda$ is found from the condition that the minimal eigenvalue of $\hat{M}^{(2)}_{ij}$ 
(understood as $2N \times 2N$ matrix) is zero in accordance with \req{Omega} and rotational invariance of the whole system.
\end{subequations}

Minimizing the energy $u_2^\Delta+u_3$ with respect to the Gaussian variables ${\mathbb R}_i$ we obtain the effective energy of the directors 
\be
\begin{split}
&\tilde{u}_D =\frac{1}{2}\bar{\mathbb D}_i{\cal M}_{ij}{\mathbb D}_j,
\\
&{\cal M}_{ij}\equiv
-\hat{M}^{(3)}_{ik}
 \left[\hat{M}^{(2)}\right]_{kl}^{-1}\hat{M}^{(3)}_{lj}
+ \frac{3}{2}
 \hat{M}^{(4)}_{ij},
\end{split}
\label{interaction}
\ee
where the inverse matrix is defined by  $\left[\hat{M}^{(2)}\right]_{ik}^{-1}\left[\hat{M}^{(2)}\right]_{kj}=\delta_{ij}\hat{\openone}$.

For the infinite system translational symmetry is restored, and the eigenvalues $\lambda$ of ${\cal M}$ are labeled by quasimomenta $q=q_x+iq_y,\ \bar{q}=q_x-iq_y$.  One easily finds
\be
\lambda_{\pm}= -\frac{|B_{20}||B_3|^2}{|B_{20}|^2- |B_2|^2} \pm \left| \frac{ \bar{B}_3^2 B_2}{|B_{20}|^2- |B_2|^2} + \frac{3\bar{B}_4}{2} \right|, \label{lambda}
\ee
where 
\be
B_n(q,\bar{q})=\sum_{\omega \neq 0} \frac{ e^{ \frac{i}{2} \left( q \bar{\omega}+ \bar{q} \omega \right)}}{\omega^n},
\ B_{20}\equiv\lim_{|q|\to 0}|B_2(q,\bar{q})|,
\label{B}
\ee
and $\omega$ labels the positions on the lattice, $\omega_{kl}\equiv a(k+l e^{i2\pi/3})$. The $B_n$ terms can be expressed  via Weierstrass elliptic functions. Doing so one finds the minimal energy configuration corresponds to the $q\to 0$ limit \cite{Note}.

Naively the $q\to 0$ state corresponds to a homogeneous configuration. However, in similar manner to ferromagnetic and ferroelectric materials, one must take into account that the boundary contribution to the energy may be of the same order as the bulk contribution \cite{Kittel1946}. Thus the actual configuration is always inhomogeneous, but smooth on scales of the lattice constant. 
 
Investigation of such configurations can be performed by going to the continuous limit of the energy \rref{higherorder}. Expanding $B_2,B_3$ to third order in $q \ll 1$ and replacing summation by integration we obtain

\begin{equation}
 u=\int d^2 r
\frac{ \partial_z \Lambda \partial_{\bar{z}}\Lambda}{4\pi}
+ \int_{r<R} \frac{ d^2 r}{\sqrt{3}a^2}
\left[\left(\frac{\Delta}{2}+{F}
\right)\partial^2_z\Delta+\cc\right]
\end{equation}
where
\begin{align}
  \label{functional}
  - \partial^2_{z\bar{z}}\Lambda &= 2\pi\rho(z,\bar{z}) \\
  \partial_{\bar{z}}{F} &= \pi\left(\frac{2}{\sqrt{3}a^2}\right)D(z,\bar{z}); \quad |D(z,\bar{z})|^2=D_*^2; \nonumber\\
  \rho(z,\bar{z}) &= \frac{2}{\sqrt{3}a^2}
  \left[ - \left(
      \partial_z\Delta+\partial_{\bar z}\bar{\Delta}
    \right) +\delta(|z|^2-R^2){\rm Re}\bar{z}\Delta \right].\nonumber
\end{align}
The field $\rho$ represents the change in vortex density due to the deformation $\Delta$, and the final term of $\rho$ describes accumulation of surface charge due to the shift of the lattice with respect to the background.

\begin{figure}[t]
  \begin{subfigure}[]{}
    \includegraphics[width=0.45\columnwidth]{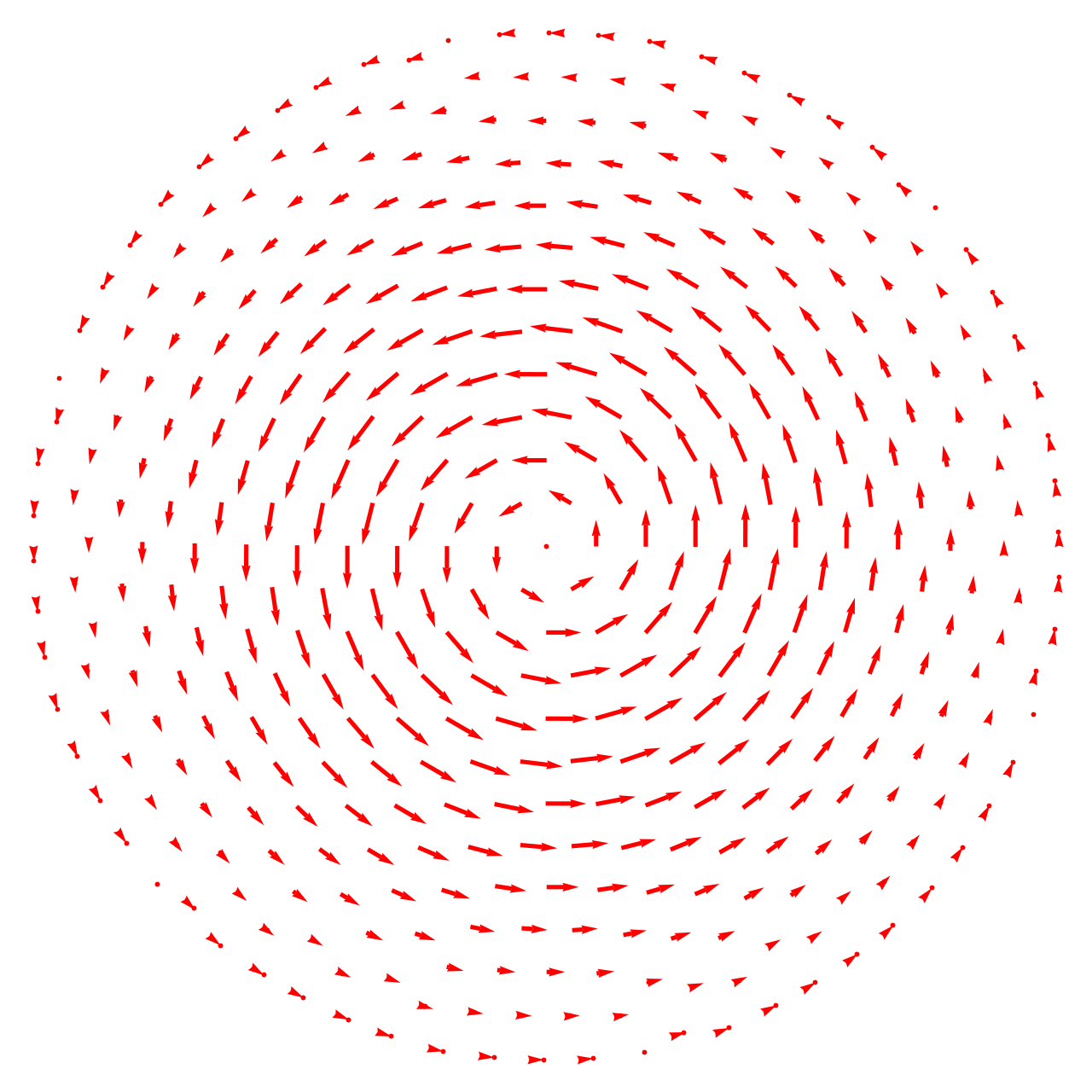}
  \end{subfigure}
    \begin{subfigure}[]{}
    \includegraphics[width=0.45\columnwidth]{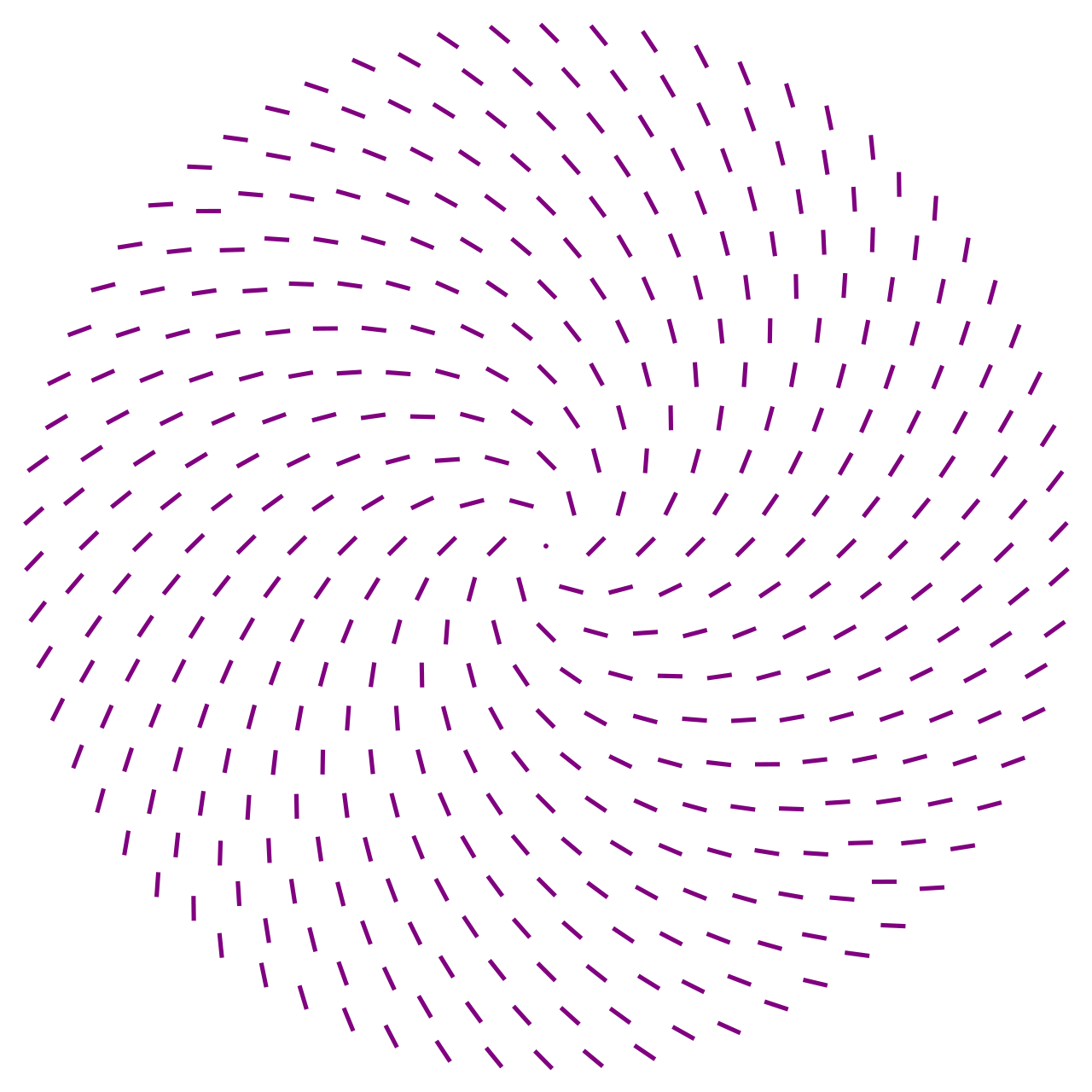}
  \end{subfigure}
  \caption{An illustration of the supervortex structure which is a minimum of the energy $u$, ignoring surface contributions [eqs. \eqref{functional}-\eqref{configuration}]. (a) the shifts, $\Delta_i$, of the vortices from their equilibrium position due to core reconstruction. (b) the configuration of $D_i$ (see Fig. \ref{fig1} for the convention used in drawing the directors).}
 \label{fig2}
\end{figure}

The functional \rref{functional} is minimized with respect to $\Delta$ and $D$ \cite{Note}. The  energetically profitable configurations should couple $D$ with $\Delta$ such that $\rho=0$. Inspection shows that having the $D$ constant throughout the system will not satisfy this property. (Due to the $B_3$ coupling, $\partial_{\bar{z}}\Delta\propto D$ and so a constant $D$ necessarily induces surface charge.) The most plausible way to avoid the surface charge is to create a "supervortex" configuration preserving the rotational symmetry of the problem,
\be
D(z,\bar{z})=D_*\frac{z}{\bar{z}}\exp[i\alpha(\sqrt{z\bar{z}})].
\label{Dtrial}
\ee
For $\alpha=0,\pi$, the reflectional symmetry is preserved as well (as we saw the surface energy term
tends to preserve both reflectional and rotational symmetry).
Substituting \req{Dtrial} in Eq.~(\ref{functional}), and minimizing with respect to $\alpha$ with the boundary condition
$\alpha(R)=\alpha_R$ we find
the solution $|\sin \alpha(r)|=\min\left(1,(R/r)^2|\sin\alpha_R|\right)$, and the resulting bulk energy
\be
u=-\left(\frac{2\pi R^2}{\sqrt{3}a^2}\right)\left(\frac{2\pi D_*}{\sqrt{3}a^2}\right)^2
\left[|\sin\alpha_R|-\frac{\sin^2\alpha_R}{2}\right].
\label{configuration}
\ee
where we have neglected logarithmic corrections in $R$ that are typical for such 2D systems. The first factor has the meaning of the number of vortices in the system, while the second is $2|\lambda_-|$ at $q=0$. 
The function $u(\alpha_R)$ has shallow minima for $\alpha=\pm\pi/2$,
i.e. for the states with broken reflectional symmetry. Fig. \ref{fig2} illustrates the configuration for $\alpha=+\pi/2$. Near the boundary,  the surface energy term (which preserves the reflectional symmetry) becomes important. It gives rise to deformation of the supervortex configurations in a small boundary layer.

{\em In conclusion}, we showed that the Abrikosov vortex lattice is
unstable with respect to non-analytic deformations of the vortex core. The
directors characterizing such deformations are themselves ordered in
``supervortex'' structures due either to surface effects or to
interactions with the shear deformations of the vortex lattice.
As a final comment, we point out that the description of static order parameter in superconductors also can be recast in a form \req{eq:3}. The validity
of condition Eq. \req{condition} should be replaced with constraints
on the applied magnetic field $H_{c2}\ln(\xi/L_T)<H<H_{c2}$, where
$L_T$ is the characteristic length determining the locality of the
Ginzburg-Landau description, $H_{c2}$ is the critical magnetic field.
Since in the vicinity of the critical temperature $\xi/L_T$ diverges,
we expect the instability of the Abrikosov lattice towards the
deformation of the vortex core to be observable in clean
superconducting films as well.

We are grateful to N.~Katz, A.~Kuklov, and E.~Zeldov for discussions of the results.  This research was supported by the United States-Israel Binational Science Foundation (BSF) grant No. 2012-134, and the Israel Science Foundation (ISF) grant No. 302/14 (O.A.), and by the Simons foundation (I.A.).

\renewcommand{\theequation}{S.\arabic{equation}}
\setcounter{equation}{0}

\renewcommand{\thefigure}{S.\arabic{figure}}
\setcounter{figure}{0}

\newpage

{\bf Supplementary material for  ``Instability of Abrikosov lattice due to nonanalytic core reconstruction of vortices in Bosonic superfluids''}
\\[1.5cm]

In this supplementary material we provide details of the calculations of $V^{(2)}_i$, $B_n(q,\bar{q})$, the  solution of the Abrikosov lattice deformation in the continuous limit of large systems. To shorten notations, in what follows we will measure distances in units of the lattice constant $a$.     

\subsection{Calculation of $\bar{V}_i^{(2)}$}
In order to calculate 
\be
\bar{V}_i^{(2)}=\bar{V}^{(2)}(z_i)=\sum_j\frac{1}{(z_i-z_j)^2},
\ee
for a finite system, $|z_i|<R$, we neglect lattice deformation near the boundary and use Poisson's summation formula:
 \be
\bar{V}^{(2)}(z) = \sum_{\vec{q}} \-\frac{ d}{dz} \int_{R}^\infty \frac{r dr}{\frac{\sqrt{3}}{2} } \int_{-\pi}^{\pi} d\theta \frac{ e^{i  \vec{q}\cdot\vec{r}}}{ z- r e^{i\theta}},
\ee
where $\frac{\vec{q}}{2\pi}=\frac{2}{\sqrt{3}}(k_1 \vec{b}_1+ k_2 \vec{b}_2)$ where $b_1=(\frac{\sqrt{3}}{2},\frac{1}{2})$, 
and $\vec{b}_2=(0,1)$, and $k_1$ and $k_2$ are integers. Here we have used the fact that on a lattice point, $z=z_i$, the integral over all space is zero (by symmetry), thus the integral over the interior domain $r<R$ may be replaced by an integral over the exterior domain $r>R$.

Next we separate the sum over $\vec{b}$, to a sum over families  using  Miller index notation defined as following.  
The facet directions of triangular Bravias lattice are conveniently described by Miller 3-index notation. 
Denoting by $\vec{a}_1=(1,0)$, $\vec{a}_2=(-\frac{1}{2},  \frac{ \sqrt{3}}{2})$, and $\vec{a}_3= (-\frac{1}{2},  \frac{ \sqrt{3}}{2})$ 
the three principal directions of the triangular lattice, a facet going through points $\vec{a}_i/l_i$ where $l_1+l_2+l_3=0$, is 
denoted by the Miller indices $(l_1,l_2,l_3)$, where $l_i$ are integers and  at least two of them are coprime.  
Equivalent crystallographic directions which correspond to $\pi/3$ rotation, are denoted by $\langle l_1, l_2, l_3 \rangle$, all of those directions
are obtained from each other by the cycling permutations of indices, or by the changing the sign of all indices. 

Therefore, each vector of the reciprocal lattice is characterized by 
$\alpha=\{k_1,k_2\}$ with fixed ratio of $k_1/k_2$, where $k_1$ and $k_2$ are coprime integers coinciding with
first two Miller indices in 3-indices notation for the facets, $k_{1,2}=l_{1,2}$. Then, each family represents a surface in $k$-space with fixed angle: 
\be
\phi_{\alpha}= \arctan \left[ \frac{1}{\sqrt{3}}\left(1+\frac{2k_2}{k_1}\right)\right],
\ee
$(0\leq \phi<\frac{\pi}{3})$, and an absolute value
 $|\vec{q}|=  q_\alpha n$ where $n$ is a positive integer, and 
\be
q_{\alpha}=\frac{2}{\sqrt{3}}\sqrt{k_1^2+k_1 k_2+k_2^2}.
\ee
Symmetry implies that each one of these families contains 6 members obtained by rotations of $\pi/3$, or
by picking up all the possible combinations from $<k_1,k_2,-k_1-k_2>$.  Thus
\be
\bar{V}^{(2)}(z) \!=\! \frac{ d}{dz}\sum_{\alpha,n} \sum_{m=0}^{5} \int_R^{\infty}\!\!\! \frac{r dr}{\frac{\sqrt{3}}{2} a^2} \int_{-\pi}^{\pi} \!\!\!\!\!d\theta \frac{ e^{i 2\pi r q_{\alpha}n \cos\left[ \phi_\alpha- \theta+ m \frac{\pi}{3} \right]}}{ z- r e^{i\theta}},
\ee
and performing the sum over $m$ we obtain
\be
\bar{V}^{(2)}(z)\!=\!\frac{2}{\sqrt{3}a^2} \sum_{\alpha,n}\frac{d}{dz}\int_R^{\infty} \!\!\! r dr \int_{-\pi}^{\pi} d\theta  \frac{ 6z^5e^{i2\pi  r q_{\alpha}n \cos(\phi_\alpha-\theta)}}{ z^6- r^{6} e^{i6\theta}}.
\ee
For $a<|z|\ll R$ we may expand to leading order in $|z|/R$ to obtain
\begin{widetext}
 \beqa
 \bar{V}^{(2)}(z)&\simeq &-\frac{60z^4}{\sqrt{3}}\sum_{\alpha,n}  \int_R^\infty \frac{dr}{r^5}\int_{-\pi}^{\pi} d\theta  e^{i 2\pi q_{\alpha}n r \cos(\phi_\alpha- \theta)-i 6\theta} \nonumber
= - \frac{60}{\sqrt{3}}\left(\frac{z}{R}\right)^4 \sum_\alpha\sum_{n=1}^\infty e^{- i 6\phi_\alpha}\frac{J_5(2\pi q_\alpha n R)}{ q_\alpha n R}.
\eeqa
\end{widetext}
Noticing that except for $\alpha=\{1,0\}$ corresponding to $\phi_\alpha=\frac{\pi}{6}$  and $\alpha=\{2,\bar{1}\}$ corresponding to $\phi_\alpha=0$, 
all other  contributions come in pairs, such that for some $0<\phi_n<\pi/6$ there is by symmetry another pair of  indices $\alpha'$ such that $\phi_\alpha'= \frac{\pi}{3}- \phi_\alpha$, and $q_\alpha=q_{\alpha'}$ we obtain Eq. [13a] with
\be
f(x)=-\frac{60}{\sqrt{3}}\sqrt{x} \sum_{\alpha,n}\delta_\alpha \cos(  6\phi_\alpha) \frac{J_5\left( 2\pi q_\alpha nx\right)}{q_\alpha n},
\ee
where the sum is over all vectors $\alpha$ such that $0\leq \phi_\alpha\leq \pi/6 $, $\delta_\alpha=1 $ if $\alpha=\{1,0\}$ or $\alpha=\{2,\bar{1}\}$ and $\delta_\alpha=2$ otherwise.  

\begin{figure}[t]
\includegraphics[width=0.9\columnwidth]{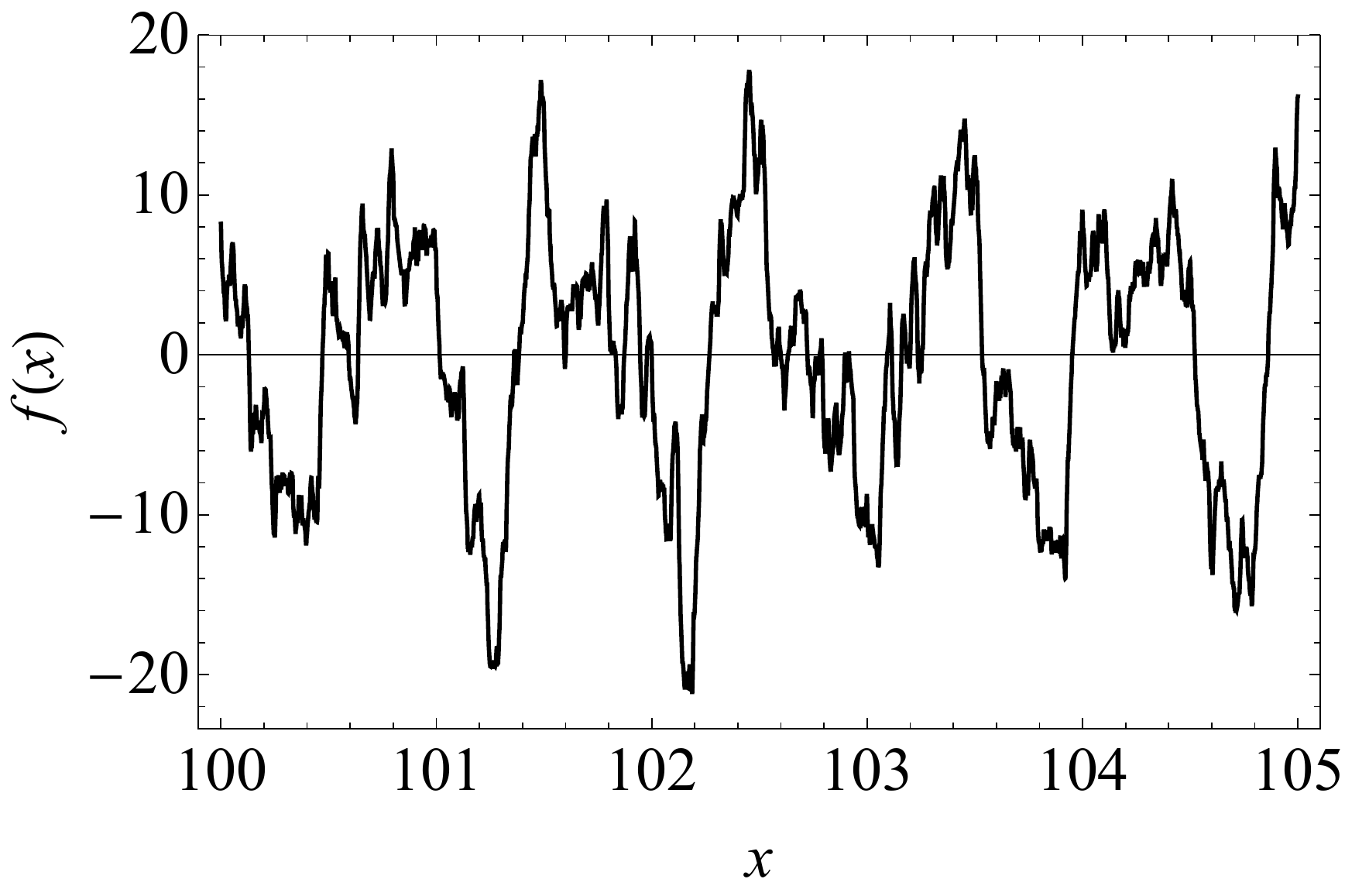}
 \caption{The function $f(x)$.
}
 \label{fig:f_x}
\end{figure}

Using the large argument asymptotic limit of the Bessel function one can sum over $n$ and express the result in the form:
\be
f(x)\simeq -\frac{120\sqrt{2}}{\sqrt{3}} \sum_{\alpha}\delta_\alpha \frac{\cos(  6\phi_\alpha)}{q_\alpha^{3/2}}\zeta\left(-\frac{1}{2},q_\alpha x\right), 
\ee
where 
\be
\zeta(\beta, \gamma)= \frac{2\Gamma(1-\beta)}{(2\pi)^{1-\beta}}\sum_{n=1}^{\infty} \frac{\sin\left(2\pi \gamma n+\beta \frac{\pi}{2}\right)}{n^{1-\beta}}
\ee
is a generalized Riemann $\zeta$-function.  A sample region of $f(x)$ is depicted in Fig. \ref{fig:f_x}.

Consider now the behavior of $\bar{V}_i^{(2)}$ near the boundary of the system, i.e. when $R-|z_i|\lesssim 1$. We focus our attention on the points $z_i$ which are close to the main facets of the triangular lattice (in real space), and ignore the shifts in the positions of the vortices that are located near the boundary (which is of order of the lattice constant). The small deviations of the boundary from a straight facet can be considered as a set of quasi random kinks. These kinks create a pseudorandom long-range fields in addition to a strong short-range field associated with the straight facets of the lattice.

In what follows we shall focus our attention on the principle facets:  $\langle 1, \bar{1}, 0 \rangle$ and $\langle 2, \bar{1}, \bar{1} \rangle$
in Miller 3-index notation.

Consider the function,
\be
\bar{V}^{(2)}(z) =- \sum_{|z_j|>R}\frac{1}{(z-z_j)^2} \label{Vz}
\ee
which is analytic in $|z|<R$ and has second order poles on the lattice points out side the system. To begin with, let us assume a large enough system and a straight facet. Then $\bar{V}^{(2)}(z)$ assumes a constant value for $z$  which lies along a straight line parallel to the facet. We denote by $\bar{V}^{\langle l_1l_2 l_3\rangle}_{(k) }$ the value of this function along lattice rows parallel the facet associated with Miller indices $(l_1l_2 l_3)$. Here $k$ denotes the row number counted from outside, i.e. $k=1$ corresponds to the boundary row, $k=2$ to the next row in, and so on. The analytic structure of $\bar{V}^{(2)}(z )$ implies that close to the $\langle 1, \bar{1}, 0 \rangle$ facet:
\be
\bar{V}^{\langle 1\bar{1}0 \rangle}_{(k)} =-\sum_{j=k}^\infty\left( \frac{\pi e^{i\frac{2\pi}{3}}}{\sin(\pi e^{i\frac{2\pi}{3}}j)}\right)^2=\eta^{\langle 1\bar{1}0 \rangle}_k e^{-2i\phi},
\ee
where $\phi$ is the angle of the facet, $e^{-2i\phi}= \frac{\bar{z}}{z}$, while
\be
 \eta^{\langle 1\bar{1}0 \rangle}_k  =-\sum_{j=k}^\infty \frac{\pi^2}{\sinh^2\left(\frac{\pi \sqrt{3}}{2}j +i \frac{\pi}{2} j\right) }.
\label{eta110}\ee
In particular, $\eta^{\langle 1\bar{1}0 \rangle}_1\simeq0.168$, $\eta^{\langle 1\bar{1}0 \rangle}_2\simeq-0.7~10^{-4}$ and $\eta^{\langle 1\bar{1}0 \rangle}_3\simeq3~ 10^{-6}$. A similar calculation for the $\langle 2, \bar{1}, \bar{1} \rangle$ facet gives $\bar{V}^{\langle  2 \bar{1} \bar{1} \rangle}_{(k)}=\eta^{\langle  2 \bar{1} \bar{1} \rangle}_{k} e^{-2i \phi}$, where
\be
 \eta^{ \langle 2 \bar{1} \bar{1}\rangle}_k  =-\sum_{j=k}^\infty \frac{\pi^2}{3\sinh^2\left(\frac{\pi }{2 \sqrt{3}}j +i \frac{\pi}{2} j\right) }.
\label{eta211}\ee 
In particular, $\eta^{ \langle 2 \bar{1} \bar{1}\rangle}_1\simeq 1.26$, $\eta^{ \langle 2 \bar{1} \bar{1}\rangle}_2\simeq -0.32$,and  $\eta^{ \langle 2 \bar{1} \bar{1}\rangle}_3\simeq 0.048$. The dependence of $\eta^{\left\langle l_1l_2l_3 \right\rangle}$ on the type of facet stems from the distance between vortices on each facet, which can vary significantly. (E.g. for the $\langle 1\b{1}0\rangle$ facet the separation is $a$, whereas for the $\left<2\b{1}\b{1}\right>$ facet it is $\sqrt{3}a$.) 

In order to generalize this result to the case where the boundary is a straight line with random kinks (generating the long-range contribution to the field), let us parameterize the boundary as:
\be
z(t)=\left\{ \begin{array}{cc} e^{-i \frac{2\pi}{3}} t + h(-t)  & \mbox{near $\langle 1, \bar{1}, 0 \rangle$  facet} \\
i t+ h(t) & \mbox{near $\langle 2, \bar{1}, \bar{1} \rangle$  facet} \end{array} \right.,
\ee
where $h(t)$ is a real function, and let us denote by $\tilde{h}(t)$ its integer part. 

Consider now the  $\langle 2, \bar{1}, \bar{1} \rangle$  facet. Here $t=(z- \bar{z})/2i$ and  therefore $z+\bar{z}= 2 h\left(i\frac{z-\bar{z}}{2}\right)$.
Thus $\tilde{h}\left(i\frac{z-\bar{z}}{2}\right)$ is analytic function in the $z$-plane with cuts parallel to the real axis. Namely,
\be
\frac{\partial}{\partial \bar{z}}\tilde{h}\left(i\frac{z-\bar{z}}{2}\right)=
\frac{i}{2} \sum_k s_k\delta\left(i\frac{z-\bar{z}}{2}-y_k\right),
\ee
where $y_k$ are the points where $\tilde{h}(y)$ jumps from value to another and $s_k= \tilde{h}(y_k+0)- \tilde{h}(y_k-0)=\pm 1$. With these definitions, the analytic function  (\ref{Vz}) can be written in the form:
\begin{widetext}
\begin{subequations}
\be
\bar{V}^{(2)}(z)\!=\!-\!\!\sum_{j=1+\tilde{h}\left( \frac{\bar{z}-z}{2i}\right)}\left(\frac{i\pi/3}{\sin\left[\frac{i \pi}{\sqrt{3}}(z+j e^{i\frac{2\pi}{3}})\right]}\right)^2- \int\frac{ d^2 z_1}{2\pi i} \frac{1}{z-z_1} \frac{\partial \tilde{h}\left( \frac{\bar{z}_1-z_1}{2i}\right)}{\partial \bar{z}_1}\left(\frac{i\pi/3}{\sin\left[\frac{i \pi}{\sqrt{3}}(z_1-1-\tilde{h}\left( \frac{\bar{z}-z}{2i}\right)\right]}\right)^2.
\ee
Here the first term accounts for the strong and short-range interaction coming from the local principal facet. The second contribution is an integral along the cuts of $\tilde{h}\left( \frac{\bar{z}_1-z_1}{2i}\right)$. As  $1/\sin^2$ is a rapidly decaying function along the line of integration, this term represents the long range contribution from pseudo random charges associated with the surface kinks.  

A similar expression is obtained for the   $\langle 1, \bar{1}, 0 \rangle$  facet:
\be
\bar{V}^{(2)}(z)\!=\!\!-\!\!\!\!\!\!\!\!\!\sum_{j=1+\tilde{h}\!\!\left( \frac{z-\bar{z}}{\sqrt{3}i}\right)}\!\left(\frac{\pi e^{i\frac{2\pi}{3}}}{\sin\left[\pi(z-j) e^{i\frac{2\pi}{3}}\right]}\right)^2\!\!\!- \int\frac{ d^2 z_1}{2\pi i} \frac{1}{z-z_1} \frac{\partial \tilde{h}\left( \frac{z_1-\bar{z}_1}{\sqrt{3}i}\right)}{\partial \bar{z}_1}\left(\frac{\pi e^{i\frac{2\pi}{3}}}{\sin\left[ \pi\left(z_1-1-\tilde{h}\left( \frac{z-\bar{z}}{\sqrt{3}i}\right)\right)e^{i\frac{2\pi}{3}}\right]}\right)^2.
\ee 
\end{subequations}
\end{widetext}

In the previous consideration we assumed that the triangular lattice is not deformed at all near the boundary. The effect of shifts 
 of the vortices from the triangular lattice to their equilibrium position could be investigated only numerically.
We found that in the finite systems  the analytic  results are modified quantitatively but not qualitatively. 

\begin{figure}[t]
\includegraphics[width=0.9\columnwidth]{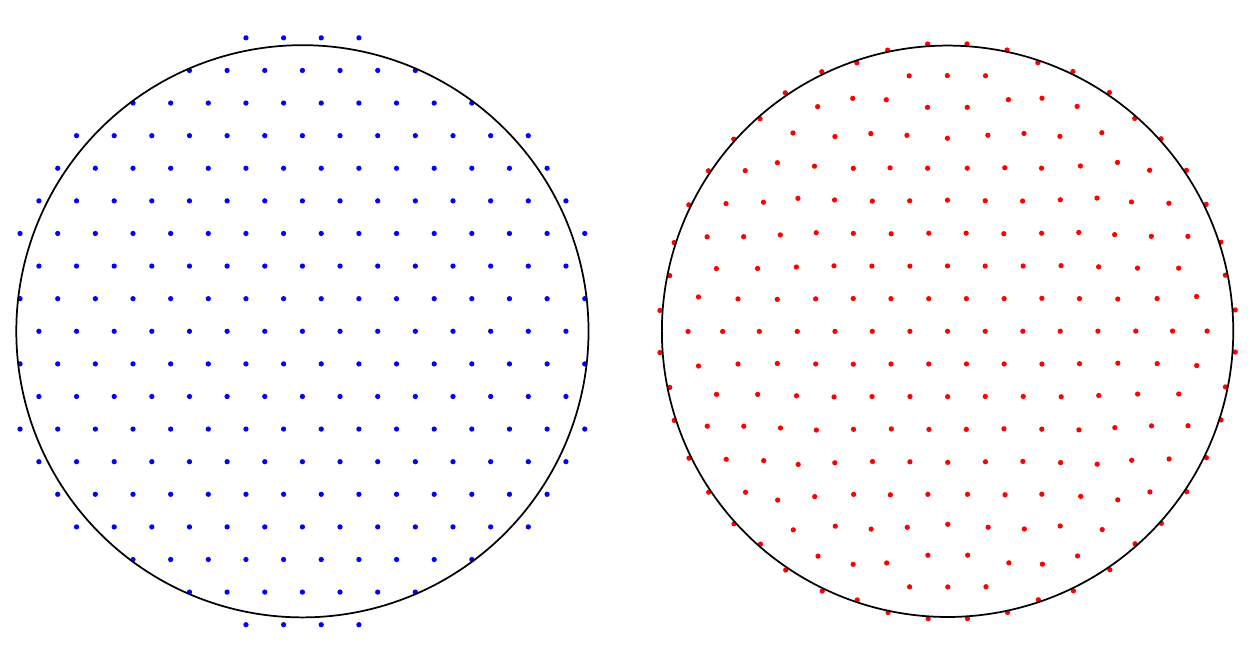}
 \caption{The vortex lattice configuration: Before relaxation (left) and after relaxation (right). The system contains 235 vortices, and the circle is only a guide to the eye. }
 \label{fig:VortexLattice}
\end{figure}

Figure \ref{fig:VortexLattice} depicts the vortex lattice structure before and after relaxation (ignoring the small effect of core reconstruction). 

\begin{figure}[t]
\includegraphics[width=0.8\columnwidth]{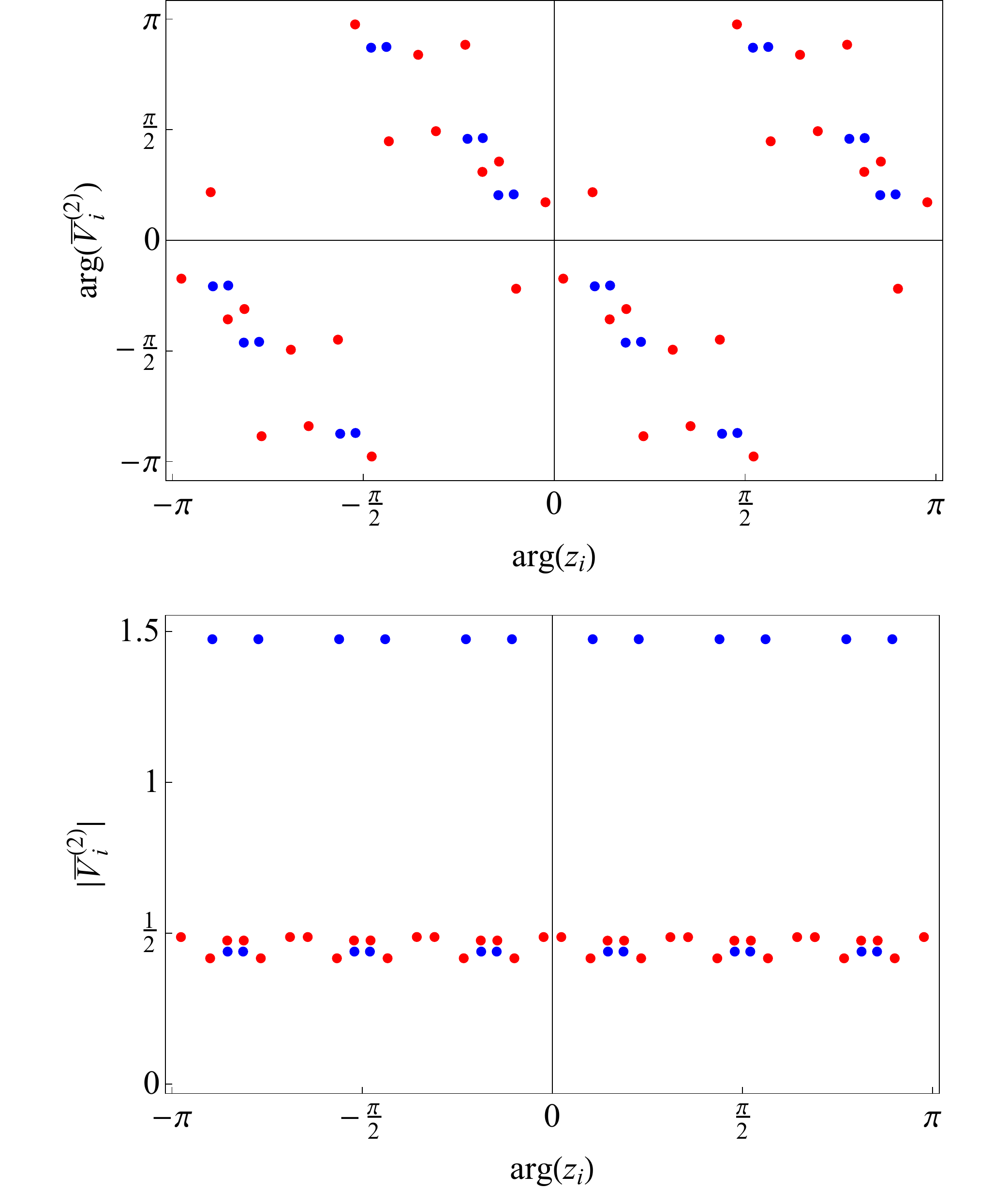}
 \caption{The argument (up) and the absolute value (down) of $V_i^{(2)}$ at the last row of the system shown in Fig.~\ref{fig:VortexLattice}. The red and blue disks represent the result for relaxed and non relaxed configurations of the vortices.}
 \label{fig:V2_numerics_surf}
\end{figure}

As is evident from the figure, the most significant result of relaxation is that boundary vortices form a circle, similarly to the situation
for point charges interacting via $1/r$ interaction, see Ref.~[14] in the main text. 
The vortices on this circle are approximately uniformly distributed with a separation of one lattice constant. 
The rest of lattice remains approximately unperturbed. The strong dependence of $\eta^{\left\langle l_1l_2l_3\right\rangle}$ on the facet stems from the effective lattice constant of each facet, 
and therefore the almost uniform lattice spacing on the boundary circle reduces the fluctuations. Fig. \ref{fig:V2_numerics_surf} shows the argument and the 
absolute value of $\bar{V}^{(2)}_i$ calculated for the outer row of the system depicted in Fig. \ref{fig:VortexLattice}. 
One can see, the angular $\bar{z}/z$ dependence of the field is robust, but the modulation of the absolute value is greatly reduced. 


\begin{figure}[t]
\includegraphics[width=0.8\columnwidth]{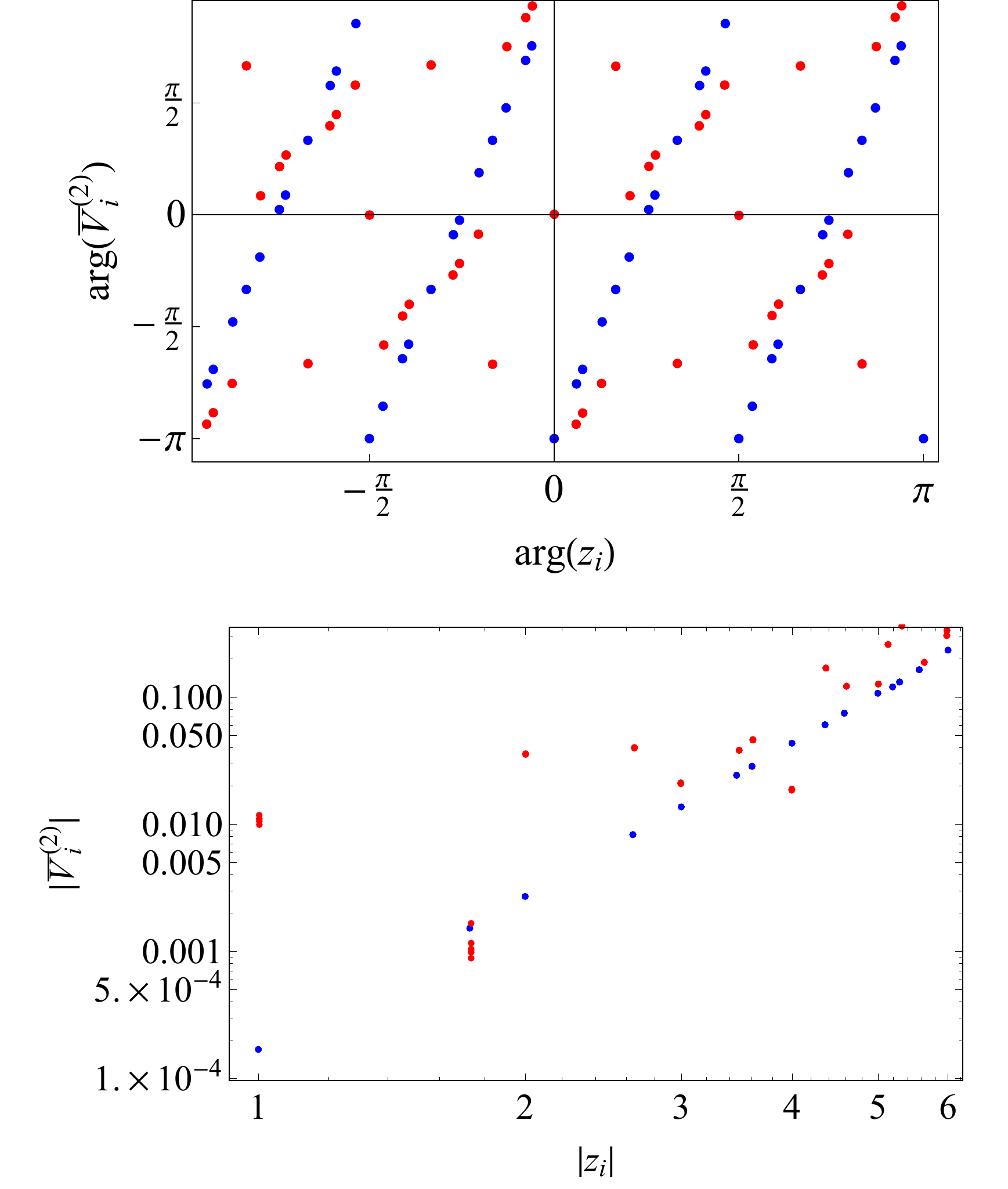}
 \caption{The behavior of $\bar{V}^{(2)}_i$ in the bulk the system shown in Fig.~\ref{fig:VortexLattice}. The upper panel shows
 the argument of $\bar{V}_i^{(2)}$  for $3<\frac{|z_i|}{a}<5$ for (where $R=7.7$). The lower panel is a log-log plot of the absolute value of $V_i^{(2)}$ for $1\leq\frac{|z_i|}{a}\leq6$. The red and blue disks represent the results for relaxed and non relaxed configurations of the vortices.}
 \label{fig:V2_numerics_bulk}
\end{figure}

Finally, the shift of the vortices due to relaxation generates
effective dipole charges near the edge of the system. The effect of
these dipoles in the bulk is smaller by an order of $a/R$ than
$V_i^{(2)}$, but they cause strong deviations near the boundary. In
upper panel of Fig. \ref{fig:V2_numerics_bulk} we show the behavior of
the argument of $\bar{V}^{(2)}_i$ as function of $\arg(z_i)$ within
the bulk. It shows that there is another contribution in addition to
the $z^4$ dependence obtained in Eq.~(13a). The
absolute value of $\bar{V}^{(2)}_i$ as function of $|z_i|$ is
presented by a log-log plot in the lower panel of the figure. The
non-relaxed lattice clearly shows the $|z|^4$ behavior, while the
results for the relaxed lattice show some deviations. However, we attribute these
bulk errors to numerical inaccuracy of the relaxation algorithm rather than to the true physical effect.

\subsection{Calculation of $B_n(q, \bar{q})$ and the eigenvalues of Eq.~ (17)}

We begin with the expression (18) given as an infinite sum,
where $\omega = n \omega_1 + m \omega_2, \omega_1 = 1, \omega_2 = e^{i\pi/3}$ define the triangular lattice. $B_n(q,\bar{q})$ inherits several properties from the lattice structure,
\begin{subequations}\label{eq:B_q_symmetry}
  \begin{align}
    \overline{B_n(q,\bar{q})} &= B_n(\bar{q}, q)   \label{B_ref};\\
    B_n(e^{i\frac{\pi}{3}}q,e^{-i\frac{\pi}{3}}\bar{q}) &= e^{-i\frac{n\pi}{3}}B_n(q,\bar{q}); \label{B_rot}\\
    B_n(q+\nu, \bar{q}+\bar{\nu}) &= B_n(q, \bar{q});\label{B_lat}
  \end{align}
\end{subequations}
where $\nu = n\nu_+ + m \nu_-, \nu_\pm = 2\pi(1 \pm \frac{i}{\sqrt{3}})$ define the reciprocal lattice. To identify the general form of $B_n$ we notice that
\beqa 
  \label{eq:B_n_deriv}
  \frac{\partial^n B_n}{\partial \bar{q}^n} &=& \left(\frac{i}{2}\right)^n \sum_{\omega \neq 0} e^{\frac{i}{2} (q\bar{\omega} + \bar{q} \omega)}\\ &=& \left(\frac{i}{2}\right)^n \left[\sum_{\omega} e^{\frac{i}{2} (q\bar{\omega} + \bar{q} \omega)} - 1\right].\nonumber
\eeqa
For any $q \neq \nu$ the infinite sum vanishes, and therefore $B_n(q, \bar{q})$ should have the form:
\begin{widetext}
\begin{equation}
  \label{eq:B_n_form}
  B_n(q, \bar{q}) = - \left(\frac{i}{2}\right)^n \frac{1}{n!}\left[\bar{q}^n + \bar{q}^{n-1} f_{1}(q) + \bar{q}^{n-2} f_{2}(q) + \cdots + f_n(q)\right],
\end{equation}
\end{widetext}
where $f_k(q)$ are analytic functions of $q$ which  are quasi-periodic on the lattice. They can be expressed in terms of the Weierstrass elliptic functions $\zeta(q)$ and $\wp(q)$, which have the following properties:
\begin{subequations}
\label{periodicity}
\beqa
\zeta(q+ \nu) &=& \zeta(q) + \frac{\bar{\nu}}{\beta};~~~~\beta= \frac{8\pi}{\sqrt{3}} \label{qpz}\\
\wp(q+\nu) &=& \wp(q);
\eeqa
\end{subequations}
where $\nu$ is any vector of the reciprocal lattice. In the limit $q \to 0$ their singular behaviour is:
\begin{subequations}
\label{aysmW}
\beqa
\zeta(q) &=&  \frac{1}{q}+ O(q^3), \\
\wp(q) &=& \frac{1}{q^2}+ O(q^4).
\eeqa
\end{subequations}
Consider first the case $n = 2$:
\be
  \label{eq:B_2-form}
  B_2(q,\bar{q}) = \frac{1}{8} \left[\bar{q}^2 + \bar{q} f_1(q) + f_0(q)\right].
\ee
In order to have property(\ref{B_lat}), the shift in $\bar{q}$  must be compensated by the quasi-periodic behavior of $\zeta(q)$, (\ref{qpz}), i.e. it can only contain the combination $\bar{q}-\beta \zeta(q)$. Thus  
\be
  \label{eq:B_2a}
  B_2(q,\bar{q}) = \frac{1}{8}\left\{\left[\bar{q}- \beta\zeta(q)\right]^2+ C_2(q)\right\},
\ee
where $C_2(q)$ is some periodic function on the lattice. This function is uniquely determined from the condition that $B_2(q,\bar{q})$ does not diverge at $|q|\to 0$. Taking into account the asymptotic behavior (\ref{aysmW}) we obtain:
\begin{equation}
  \label{eq:B_2-final}
  B_2 = \frac{1}{8} \left\{ \left[\bar{q} - \beta \zeta(q)\right]^2 -\beta^2\wp(q)\right\}.
\end{equation}
From here it follows that the small $q$ asymptotic behavior is:
\be
B_2(q,\bar q) = -\frac{2\pi}{\sqrt{3}} \frac{\bar{q}}{q}+\frac{1}{8} \bar{q}^2;~~~|q|\ll 1.
\ee

\begin{figure}[t]
\includegraphics[width=0.8\columnwidth]{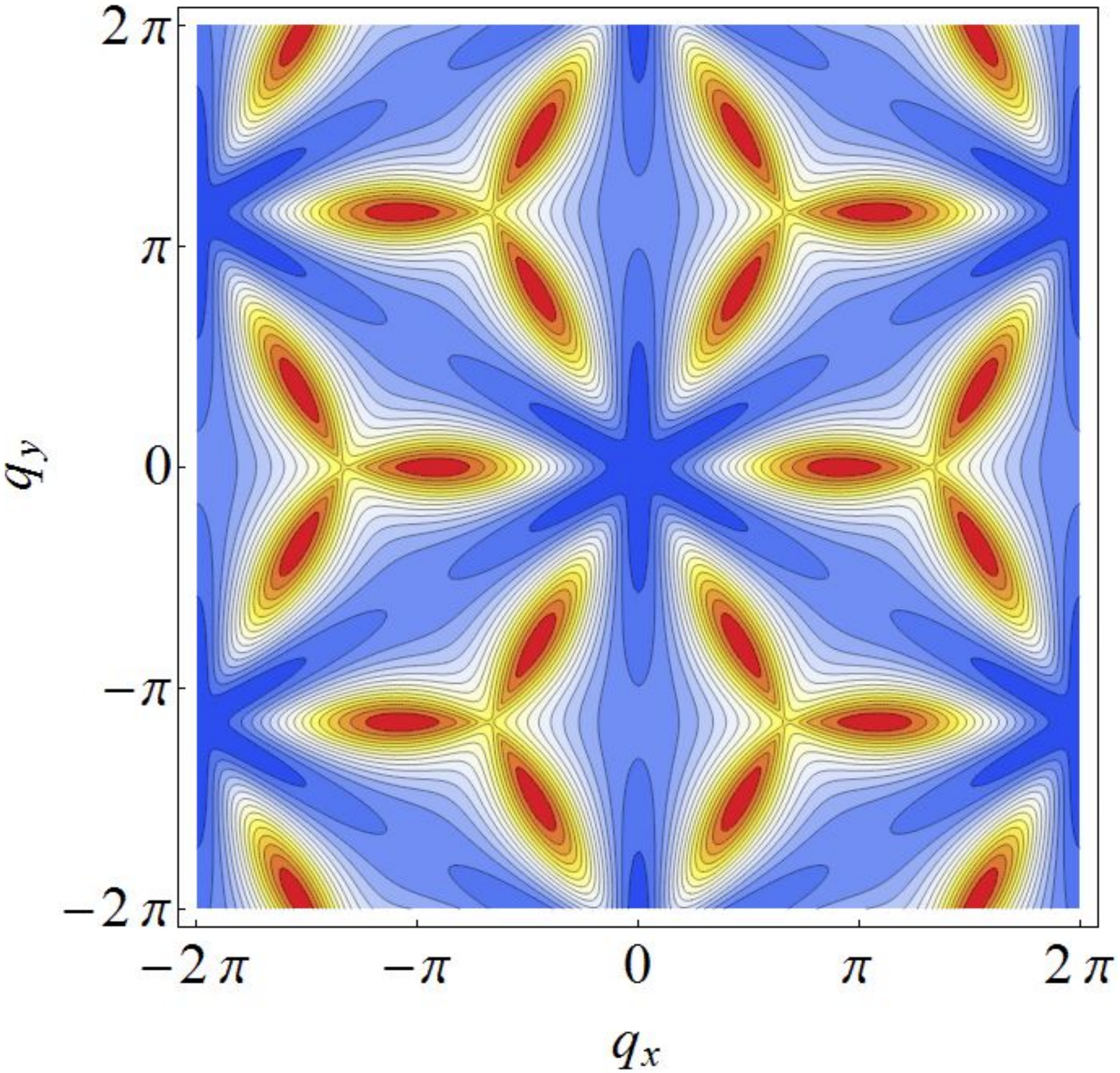}
 \caption{A density plot of the eigenvalue $\lambda$ in $q$ space. Blue colures denotes small values while red color represents high value of $\lambda$.
 The hexagon denotes the first Brillouin zone. A minimum at the $\mathcal{K}$ would imply tripling of the lattice constant, while a minimum at the  $\mathcal{M}$-point will generate stipe phase. However, the global minimum, is at the $\Gamma$ -point, similar to ferroelectric systems.
}
 \label{fig:energy-analytic}
\end{figure}

 The calculation of the $B_3(q,\bar q)$ and $B_4(q,\bar q)$ follows a similar procedure. For instance from (\ref{eq:B_n_form}) it follows that 
\be
  \label{eq:B_3a}
  B_3(q,\bar{q}) = \frac{i}{48}\left\{\left[\bar{q}-\beta \zeta(q)\right]^3+ C_3(q)\right\},
\ee 
where again $C_3(q)$ is periodic function which removes the singularities of the first term. The result is:
\begin{equation}
  \label{eq:B_3-final}
  B_3 = \frac{i}{48} \left[ \tilde{\zeta}^3 -3\beta^2\tilde{\zeta}\wp(q)+\beta^3 \wp'(q)\right],
\end{equation}
where
\be 
\tilde{\zeta}=\bar{q}-\beta \zeta(q).
\ee
The asymptotic behavior in this case is:
\begin{equation}
  \label{eq:B_3a-asymptotic}
  B_3(q,\bar{q}) =- \frac{i\pi}{2\sqrt{3}} \frac{\bar{q}^2}{q}+ \frac{i\bar{q}^3}{48}; ~~~~|q| \ll 1.
  \end{equation}
Finally,  for $n=4$ we have:
\begin{equation}
  \label{eq:B_4-final}
  B_4 \!=\! \frac{-1}{384} \left[ \tilde{\zeta}^4 -6 \beta^2 \tilde{\zeta}^2\wp(q)+ 4 \beta^3 \tilde{\zeta} \wp'(q)-3\beta^4 \wp^2(q)\right].
\end{equation} 
Using the above expressions for $B_n$ and Eq.~(17) for the eigenvalue $\lambda_-$ one obtains the energy of the lattice per unit cell, as function of $q$. A density plot of this energy is shown in Fig. \ref{fig:energy-analytic}.
 
\subsection{The energy and configuration of large systems}
 
Minimizing the functional (19) with respect to $\Delta$ within the bulk we obtain
\be
\frac{\partial^2 \Delta}{\partial z^2}=- \frac{\partial^2 F}{\partial z^2}-2 \frac{\partial \Lambda}{\partial z}; ~~~~~ \frac{\partial^2 \bar{\Delta}}{\partial \bar{z}^2}=- \frac{\partial^2 \bar{F}}{\partial \bar{z}^2}-2 \frac{\partial \Lambda}{\partial \bar{z}}. \label{eq:D2}
\ee
The solution of the first of these equations is:
\be
 \frac{\partial \Delta}{\partial z}=- \frac{\partial F}{\partial z}-2\Lambda+ \bar{\lambda}(\bar{z}),
\ee
where $\bar{\lambda}(\bar{z})$ is the integration constant. However $\partial \Delta/\partial z$ is a scalar and therefore $\bar{\lambda}(\bar{z})$ can only be a constant. This constant represents a rigid rotation of the system and may be set to be zero.  Thus
\be
 \frac{\partial \Delta}{\partial z}=- \frac{\partial F}{\partial z}-2 \Lambda; ~~~~~ \frac{\partial \bar{\Delta}}{\partial \bar{z}}=- \frac{\partial \bar{F}}{\partial \bar{z}}-2 \Lambda. \label{eq:D1}
\ee 
Thus within the bulk
\be
\rho= \frac{2}{\sqrt{3}}\left(\frac{\partial F}{\partial z}+\frac{\partial \bar{F}}{\partial \bar{z}}+4\Lambda\right),
\ee
and substituting the second equation of (20) we obtain:
\be
\frac{1}{2\pi} \frac{\partial^2 \Lambda}{\partial z \partial\bar{z}}=  \frac{\partial F}{\partial z}+\frac{\partial \bar{F}}{\partial \bar{z}}+4\Lambda.
\label{eq:Lambda}
\ee
This equation can be solved by iterations. In the first approximation one neglects the gradient term of $\Lambda$ to obtain:
\be
\Lambda_0= -\frac{1}{4} \left(\frac{\partial F}{\partial z}+\frac{\partial \bar{F}}{\partial \bar{z}}\right),
\ee
while next iteration gives
\be
\Lambda_1= -\frac{1}{4} \left(\frac{\partial F}{\partial z}+\frac{\partial \bar{F}}{\partial \bar{z}}\right)+\frac{1}{8\pi} \frac{\partial^2 \Lambda_0}{\partial z \partial\bar{z}}. \label{eq:Lambda1}
\ee 
Substituting the solution  (\ref{eq:D2}) in the expression for  the energy $u$  and ignoring the surface term in $\rho$ (as will be justified later) we have
\beqa
u&=& \int \frac{d^2z}{\sqrt{3}} \left[  \left(\frac{1}{2}\Delta+ F\right) \left(-2\frac{\partial \Lambda}{\partial z}- \frac{\partial^2 F}{\partial z^2}\right) + C.c\right]\nonumber \\
&+& \frac{1}{8\pi}\frac{\partial \Lambda}{\partial z}\frac{\partial\Lambda}{\partial\bar{z}}.
\eeqa 
Now integrating by parts and substituting (\ref{eq:D1}) we obtain
\be
u= \!\!\!\int\! \frac{d^2z}{\sqrt{3}} \frac{1}{2}\left[\left(\frac{\partial F}{\partial z}\right)^2\!+\!\left(\frac{\partial \bar{F}}{\partial \bar{z}}\right)^2\!\!\!-\!8\Lambda^2 \right]\!+\! \frac{1}{8\pi}\frac{\partial \Lambda}{\partial z}\frac{\partial\Lambda}{\partial\bar{z}}.
\ee 
Finally substituting the leading order approximation for $\Lambda$ (\ref{eq:Lambda1}) we obtain:
\be
u= \!\!\!\int\! \frac{d^2z}{\sqrt{3}} \frac{1}{4}\left(\frac{\partial F}{\partial z}-\frac{\partial \bar{F}}{\partial \bar{z}}\right)^2\!\!+\! \frac{1}{8\pi}\frac{\partial \Lambda_0}{\partial z}\frac{\partial\Lambda_0}{\partial\bar{z}} \label{eq:epsilon-final}.
\ee

To proceed further we assume that the directors from a cylindrically symmetric configuration (21),
where $\alpha(r)$ is some arbitrary function (to be fixed by minimization) of the radius $r=\sqrt{z\bar{z}}$. Substituting in the third equation of (20) we obtain 
\be
F(z,\bar{z})= -\frac{4\pi}{\sqrt{3}} D_* z \int_r^R\frac{dr_1}{r_1} \exp[i \alpha(r_1)],
\ee 
where we took into account that the force $F$ vanishes on the boundary $r=R$.
From here it follows
\be
\frac{\partial F(z,\bar{z})}{\partial z}= -\frac{2\pi}{\sqrt{3}} \frac{D_*}{r} \frac{d}{dr} \left\{r^2\int_r^R\frac{dr_1}{r_1} \exp[i \alpha(r_1)]\right\}.
\ee 
Now to find $\alpha(r)$ one should substitute this expression in (\ref{eq:epsilon-final})
and vary with respect to $\alpha(r)$. To first approximation we can neglect the higher gradient term and take only the first term, namely
\be
u= -\frac{(2\pi)^3}{3\sqrt{3}} D_{*}^{2}\int_0^R\! \frac{dr}{r} \left( \frac{dY(r)}{dr} \right)^2,
\ee 
where
\be
Y(r)=r^2 \int_r^R \frac{dr_1}{r_1} \sin[\alpha(r_1)].\label{Yint}
\ee
Minima appear either for $\frac{\delta Y}{\delta \alpha}=0$, giving $\alpha=\pm \frac{\pi}{2}$, or for
\be
\frac{d}{dr} \left(\frac{1}{r} \frac{d Y}{dr} \right)=0,
\ee
which gives
\be
Y(r)= \frac{R^2-r^2}{2} \sin[\alpha(R)],
\ee
implying that
\be
\sin[\alpha(r)] = \sin[\alpha(R)] \frac{R^2}{r^2}.
\ee
This solution clearly holds when $r>R^*$ for some $R^*<R$ and it should match the other solution at smaller values of $r$. Matching at $r=R^*$ we have
\be
1 = \left|\sin[\alpha(R)]\right| \frac{R^2}{R^{*2}}. \label{eq:matching}
\ee 
In order to  find $R^*$ we should minimize the total energy of the system which includes also the surface energy due to the interaction of the directors with the field $V^{(2)}_i$ which is very strong near the boundary: 
\be
u_s=-\Re\sum_i \bar{V}_i^{(2)}D_i. \label{uf1}
\ee
The orientation of the  directors at the most outer rows of the system is dictated by $V^{(2)}_i,$ however this field at inner rows is  much smaller [see Eqs.~(\ref{eta110}) and (\ref{eta211})]. In order to take it into account we substitute $V_i^{(2)}\approx \eta \frac{\bar{z}_i}{z_i}$
where $\eta\ll 1$ is a constant, the ansatz for $D_i$, and we approximate the sum (\ref{uf1}) by an integral:
\be
u_s= -\eta\frac{8\pi^{2}}{\sqrt{3}}D_* R \cos[\alpha(R)].
\ee
To calculate the bulk energy we first perform the integral (\ref{Yint}) to obtain $Y(r)$ at $r<R^*$ and find
\be
Y(r)=\sigma r^2\ln\left(\frac{R^*}{r}\right)+ \frac{r^2}{2}\left(\sigma-\sin[\alpha(R)]\right),
\ee
where $\sigma= \mbox{ sign}[ \sin[\alpha(R)]]$ and therefore
\be
\frac{dY(r)}{dr}=\left\{ \begin{array}{cc} 2\sigma r\ln \left(\frac{R^*}{r}\right)-r\sigma|\sin[\alpha(R)]| & r<R^* \\
-r\sin[\alpha(R)] & r> R^* \end{array} \right. .
\ee
Substituting in $u$ we obtain
\begin{widetext}
\be
u= -\frac{(2\pi)^3}{3\sqrt{3}}D_*^2R^2\left[ \sin^2\alpha\left(\frac{1}{2}-4\int_0^1 d\eta \eta \ln\frac{1}{\eta}\right)+ 4|\sin[\alpha(R)]|\int_0^1 d\eta \eta \ln^2\eta\right].
\ee
Both integrals in this expression equal $1/4$ and therefore the total energy (bulk and surface) is
\be
u+u_s= -\frac{(2\pi)^3}{3\sqrt{3}}D_{*}^2R^2 \left[|\sin[\alpha(R)]|-\frac{1}{2}\sin^2[\alpha(R)]\right] -\eta \frac{8\pi^2}{\sqrt{3}} D_* R \cos[\alpha(R)].\label{eq:final-energy}
\ee
\end{widetext}

This expression has two minima within the range $|\alpha(R)| \leq \frac{\pi}{2}$. For large $R$, they are given by $\alpha(R)= \pm\frac{\pi}{2}\mp (2\gamma)^{1/3}$, where  $\gamma= 3\eta/(\pi D_* R)$. Substituting this result in Eq. \eqref{eq:matching} we find:
\be
R-R^*=\frac{1}{4} (2 \gamma)^{2/3}R=\left(\frac{3\eta}{4\pi D_*}\right)^{2/3} R^{1/3}.
\ee


\begin{thebibliography}{99}

\bibitem{Abrikosov1957a}
A.A.~Abrikosov, J. Phys. Chem. Solids, {\bf 2},199 (1957).

\bibitem{Abrikosov1957}
A.A.~Abrikosov, Zh. Eksp. i Teor. Fiz.  {\bf 32}, 1442 (1957) [ Sov. Phys. JETP {\bf 5}, 1174 (1957)];  

\bibitem{Tkachenko1966}
V.K.~Tkachenko,  Zh. Eksp. Teor. Fiz. {\bf 49}, 1875 (1965) [ Sov. Phys. JETP, {\bf 23},1049 (1966)].

\bibitem{Tkachenko1966a}
V.K. Tkachenko, Zh. Eksp. Teor. Fiz. {\bf 49}, 1875
(1965) [ Sov. Phys. JETP, {\bf22}, 1282 (1966)].

\bibitem{Andereck1980}
C.~D. Andereck, J.~Chalups, and W.~I. Glaberson,Phys. Rev. Lett. {\bf 44}, 33 (1980).

\bibitem{Andereck1982}
C.~David Andereck and W.I.~Glaberson, J.  Low Temp. Phys. {\bf 48}, 257 (1982).

\bibitem{Coddington2003}
I.~Coddington, P.~Engels, V.~Schweikhard, and E.~A. Cornell,  Phys. Rev. Lett. {\bf 91} 100402 (2003).

\bibitem{Popov1973}
V.N.~Popov, Zh. Eksp. Teor. Fiz. {\bf 64}, 672 (1972) [Sov. Phys. JETP, {\bf 37},341 (1973)].

\bibitem{Popov2001}
V.N. Popov. {\em Functional integrals in quantum field theory and statistical
  physics}, Volume~8. (Kluwer Academic, 2001).

\bibitem{Ambegaokar1980}
V.~Ambegaokar, B.I.~Halperin, D.R.~Nelson, and E.D.~Siggia, Phys. Rev. B, {\bf21}, 1806 (1980).

\bibitem{Klein2014}
A.~Klein, I.L.~Aleiner, and O.~Agam, Annals of Physics, {\bf 346}, 195 (2014).

\bibitem{Radzihovsky15} L. Radzihovsky,  Phys. Rev. Lett. {\bf 115}, 247801 (2015). 

\bibitem{Stauffer1968} D.~Stauffer and A.L.~Fetter,  Phys. Rev. {\bf 168}, 156 (1968).

\bibitem{Koulakov1998} A.A.~Koulakov and B.I.~Shklovskii, Phys. Rev. B, {\bf 57}, 2352 (1998).

\bibitem{Note} See Supplemental Material for details of the calculation.

\bibitem{Kittel1946}
C.~Kittel, Phys. Rev. {\bf 70}, 965, (1946).
\end{thebibliography}
\end{document}